\documentclass{blois}
\usepackage{graphicx}
\usepackage{lipsum}
\usepackage{cancel}
\usepackage{slashed}
\usepackage{soul}
\usepackage{mathtools}
\usepackage{bm}
\usepackage{float}
\usepackage{caption}
\usepackage{cite}
\usepackage{amsmath, amssymb}
\usepackage{bbold}
\usepackage[all]{hypcap}
\usepackage{tikz}
\usepackage{tikz-feynman}
\usepackage{pgfplots}
\usepackage{bbm}
\usepackage{xcolor}
\usepackage[font=small,labelfont=bf]{caption}
\usepackage{bbm}
\usepackage{color, colortbl}
\usepackage{subcaption}
\usepackage[font=footnotesize,labelfont=bf]{caption}
\usepackage{array}
\usepackage[export]{adjustbox}
\allowdisplaybreaks

\def\be{\begin{equation}}
\def\ee{\end{equation}}
\def\bea{\begin{eqnarray}}

\def\LambdaHC{\Lambda_{\rm HC}}
\def\eea{\end{eqnarray}}

\def\heavy{{[3]}}
\def\light{{[12]}}

\def\cH{\mathcal{H}}
\def\cG{\mathcal{G}}

\def\modH{h}

\newcommand{\Photo}{}

\begin{document}
\vspace*{4cm}
\title{Flavour Non-Universality and Higgs Compositeness}

\author{Marko Pesut}

\address{Physik-Institut, Universität Zürich, Winterthurerstrasse 190\\
 CH-8057 Zürich, Switzerland}

\maketitle\abstracts{
We present a flavour non-universal extension of the Standard Model combined with the idea of Higgs compositeness~\cite{Covone:2024elw}. At the TeV scale, the gauge groups $SU(2)_R$ and $U(1)_{B-L}$ are assumed to act in a non-universal manner on light- and third-generation fermions, while the Higgs emerges as a pseudo Nambu-Goldstone boson of the spontaneous global symmetry breaking  $Sp(4)\to SU(2)_L\times SU(2)_R^{[3]}$, attributed to new strong dynamics. We discuss how the radiatively generated Higgs potential has the ingredients to realize the unavoidable tuning necessary to separate electroweak and composite scales. In particular, the flavoured gauge bosons resulting from the deconstruction must lie in the vicinity of the TeV scale, thereby providing interesting  phenomenology that can be probed at near future colliders. }

\section{Introduction}
The Standard Model (SM) faces two major theoretical problems related to the Higgs sector: the flavour puzzle and the electroweak (EW) hierarchy problem. The flavour puzzle arises from the unexplained pattern and hierarchy of Yukawa couplings spanning several orders of magnitude, while the EW hierarchy problem is related to the instability of the Higgs potential under quantum corrections from heavy dynamics. Traditional model building attempts~\cite{DAmbrosio:2002vsn} often separate these two issues, with the latter being solved by flavour-blind new physics (NP) close to the EW scale and flavour-full dynamics generating Yukawa hierarchies at much higher energies. However, the absence of direct evidence of flavour-blind NP near the electroweak scale in current experimental searches suggests exploring alternative approaches to model building~\cite{Davighi:2023iks,Allwicher:2023shc,Davighi:2022bqf, Lizana:2024jby,Davighi:2023xqn,Davighi:2023evx} that would address \emph{simultaneously} the two problems. An interesting class of models~\cite{Barbieri:2023qpf} is the one featuring flavour non-universal NP, mainly coupled to third-generation fermions and the Higgs, at the TeV scale. This structure relaxes experimental constraints by implementing accidental $U(2)^n$ flavour symmetries~\cite{Barbieri:2011ci,Barbieri:2012uh,Isidori:2012ts,Redi:2012uj} and offers an interesting path to address simultaneously these two mysteries of the SM.\\ \\
In this paper, we combine the hypothesis of a composite Higgs, in the spirit of the minimal composite Higgs model originally proposed in~\cite{Contino:2003ve,Agashe:2004rs}, with the idea of flavour deconstruction, along the lines developed recently in~\cite{Davighi:2023iks,Barbieri:2023qpf}. As we will demonstrate, combining these two hypotheses provides significant benefits, as it allows new physics coupled dominantly to the third generation and responsible for the stability of the Higgs sector to lie in the few TeV domain \footnote{The recent analysis~\cite{Glioti:2024hye} underlines the key role played by global flavour symmetries to allow for a smaller scale separation between elementary and composite dynamics.}. Similar models that combine Higgs compositeness with solving the flavour puzzle are proposed in {\em e.g.}~\cite{Lizana:2024jby,Fuentes-Martin:2020bnh,Chung:2023gcm}.

\section{Main Features of the Model}
In the spirit of composite Higgs models~\cite{Panico:2015jxa,Giudice:2007fh}, we envisage a new strongly coupled gauge group $G_{\text{HC}}$ giving rise to the spontaneous breaking of a global symmetry $\cG \equiv Sp(4)$ down to a (fully gauged) subgroup \footnote{The upper index on a gauge group factor denotes a flavour non-universal gauge symmetry acting only on the given family of chiral SM-like fermions.} $\cH \equiv SU(2)_L \times SU(2)_R^{\heavy}$. This transition happens dynamically at a scale $\Lambda_{\rm HC} \equiv 8\pi F \gg v_{\text{EW}}$ and delivers a pseudo-Nambu-Goldstone boson (pNGB) Higgs transforming as a bidoublet of the unbroken subgroup. In a second symmetry breaking step, the deconstructed EW symmetry is broken down to the SM,
\be
SU(3)_c \times  SU(2)_L \times SU(2)_R^{\heavy}  \times U(1)_{B-L}^\heavy \times U(1)_Y^\light \xrightarrow[\langle \Omega_f \rangle]{\langle \Sigma_R \rangle} SU(3)_c \times SU(2)_L \times U(1)_Y\, ,
\label{decew}
\ee
by scalar link fields $\Sigma_R$ and $\Omega_f$ acquiring vacuum expectation values (VEVs), and thereby giving masses to two $Z^\prime$ and a pair of $W_R^\pm$ flavoured gauge bosons. These scalars are assumed to be elementary for simplicity, but one could envisage a more complete model in which they are also composite. The matter content of the model can be decomposed into two main sectors: the elementary and the composite one. The elementary fields and their transformation properties are shown in Table~\ref{tab:Matter_Content}:

\begin{table}[H]
\begin{center}{\small
\begin{tabular}{|c|c||cc|cc|}
\hline
 \multicolumn{2}{|c||}{Elementary fields} &  $U(1)_{B-L}^\heavy$ & $U(1)_Y^\light$ & $SU(2)_L$ & $SU(2)_R^\heavy$ \\
\hline
chiral  & $q_L^\light$  & $0$ & $1/6$ & $\bf{2}$ & $\bf{1}$  \\
light quarks 
        & $u_R^\light$  & $0$ & $2/3$ & $\bf{1}$ & $\bf{1}$  \\ 
        & $d_R^\light$  & $0$ & $-1/3$ & $\bf{1}$ & $\bf{1}$  \\ 
\hline
chiral  & $q_L^\heavy$  & $1/6$ & $0$ & $\bf{2}$ & $\bf{1}$  \\
3$^{\rm rd}$ gen. quarks
        & $q_R^\heavy$  & $1/6$ & $0$ & $\bf{1}$ & $\bf{2}$  \\ \hline\hline
vector-like     
    & $F^q_L$  & $1/6$ & $0$ & $\bf{2}$ & $\bf{1}$  \\
quarks    
    & $F^q_R$  & $0$ & $1/6$ & $\bf{1}$ & $\bf{2}$  \\
\hline \hline
scalar & $\Sigma_R$  & $0$ & $ 1/2$ & $\bf{1}$ & $\bf{2}$ \\
link fields & $\Omega_q$  & $-1/6$ & $1/6$& $\bf{1}$ & $\bf{1}$ \\
 & $\Omega_\ell$  & $1/2$ & $-1/2$ & $\bf{1}$ & $\bf{1}$ \\
\hline
\end{tabular} 
}
\end{center}
\caption{Matter content of the elementary sector. For simplicity, among the fermions 
only the quarks are shown. 
\label{tab:Matter_Content}}
\end{table}
\noindent
The composite sector is well defined only below the scale $\Lambda_{\rm HC}$: it describes bound states of fields charged under the hypercolour gauge group which, by construction, form complete representations of $\cH$.  
Among them, the states playing a key role in the construction are:
\begin{itemize}
\item {\bf The Higgs field}, that will itself trigger EW symmetry breaking upon acquiring its own vacuum expectation value.
\item {\bf The lightest vector resonances},
namely the lightest composite spin-1 states transforming under $\cH$ as the corresponding gauge bosons. Their masses, $M_\rho$, are assumed to be below the scale $\LambdaHC$.

\item {\bf The top partners}, namely composite spin-$\frac{1}{2}$ states with masses $M_T$, that couple to the top quark and are ultimately responsible for generating its Yukawa coupling. Other fermionic resonances have masses indicated by $M_f$.
\end{itemize}
The flavour hierarchies are generated by the following steps:
\begin{itemize}
\item[I.] At high energies ($\gtrsim 100 \, \text{TeV}$), flavour structures are seeded by interactions between chiral fermions and operators in the strong sector. Non-universal gauge symmetries ensure that only third-generation fermions couple directly to the strong sector, while vector-like (VL) fermions, with masses $M_F$, mediate couplings to light chiral fermions.

\item[II.] Below $100 \, \text{TeV}$, integrating out VL fermions generates higher-dimensional operators. Once link fields acquire their VEVs, these operators introduce suppression factors $\epsilon_R$ and $\epsilon_L$, essential for realistic light Yukawa hierarchies:
\be
\epsilon_R \sim \frac{\langle \Sigma_R \rangle}{M_F}, \quad 
\epsilon_L \sim \frac{\langle \Omega_f \rangle}{M_F} \Longrightarrow Y_{u,d,e} \sim \begin{pmatrix}
            \epsilon_R & \epsilon_L \\
            \epsilon_R \epsilon_L & 1
        \end{pmatrix} \, .
\ee

\item[III.] Below $\Lambda_{\rm HC} \sim 10 \, \text{TeV}$, the strong sector undergoes global symmetry breaking $\mathcal{G} \to \mathcal{H}$, producing the SM-like Higgs as a pNGB and composite top partners.

\item[IV.] At scales $\lesssim F \sim 1 \, \text{TeV}$, integrating out the top partners results in an effective SM-like Yukawa interaction and a radiatively generated Higgs potential. The latter arises from explicit symmetry-breaking effects, such as the coupling of chiral elementary third-family fermions to composite resonances (partial compositeness \cite{Kaplan:1991dc}), the gauging of $\mathcal{H}$, and the VEV of $\Sigma_R$.
\end{itemize}

\section{Higgs Potential}
The effects of the strong dynamics on the fermion and gauge sectors are parametrized by momentum-dependent form factors. Due to the flavour non-universal gauge structure, the mechanism of partial compositeness only allows direct coupling of elementary third-family fields to composite resonances of the strong sector, thereby generating explicit breaking of the global symmetry $\cG$, and, consequently, insertions of trigonometric functions of the Higgs field. Similarly, the gauging of $\cH \subset \cG$ as well as the VEV of $\Sigma_R$ break explicitly $\cG$ and contribute to radiatively generating the Higgs potential. \\ \\ Using the Coleman-Weinberg approach, we re-sumed all the insertions of the Higgs field at one-loop and computed their contributions to the potential.
At leading order, the potential assumes the following general decomposition:
\be
V(h) = \Delta V_f(h) +\Delta V_A(h) \approx 
 c_0 -  c_1 \sin^2 \left(\frac{\modH}{2F}\right) + c_2 \sin^4 \left(\frac{\modH}{2F}\right)\,,
\label{eq:Vh}
\ee
where $\Delta V_f(h) $ and $\Delta V_A(h) $ indicate contributions originating from the fermion sector and from the gauging of $\cH$ (including the effects of $\Sigma_R$), respectively. The physical conditions imposed on $c_{1,2}$ in order to recover the SM Higgs potential at leading order are:
\begin{equation}
\left.\frac{c_1}{F^4}\right|_{\text {phys. }}=\frac{m_h^2}{F^2} \quad \text { and }\left.\quad \frac{c_2}{F^4}\right|_{\text {phys. }}=\frac{2 m_h^2}{v^2} \approx \frac{1}{2} .
\label{phcond}
\end{equation}
To obtain explicit expressions for $c_{1,2}$ in terms of model parameters, we assume a simple functional form of the form factors that ensures a convergence of the corresponding integrals, such that the logarithms in the Coleman-Weinberg one-loop potential are well approximated by their expansion to the first or second order (according to the parametric dependence on the couplings). Under these assumptions, we find the following expressions:
\be
\frac{c_1}{F^4} = \frac{N_c}{8\pi^2}\left[(\lambda^t_R)^2\kappa^t_{R}-(\lambda^t_L)^2\kappa^t_{L}\right] \frac{ M^2_{f} }{F^2}+ \frac{N_c y_t^2}{4\pi^2}   \frac{M_T^2}{F^2} 
-\frac{9 g_R^2}{32\pi^2}  \left(1 -\frac{g_R^2v^2_\Sigma }{2 M_\rho^2} \right) \frac{M_\rho^2}{F^2}
+\mathcal{O}(g_Lg_R,g_L^2)\,,
\label{c1eq}
\ee
where $\lambda^t_{L,R}$ and $\kappa^t_{L,R}$ are couplings of third-family fields to composite resonances, $g_R$ is the gauge coupling of $SU(2)_R^\heavy$ and $v_\Sigma = \langle \Sigma_R \rangle$. For $c_2$, we get:
\be \label{eq:c2formula}
\frac{c_2}{F^4} =  \frac{N_c y_t^2}{4\pi^2} \frac{ M_T^2 }{F^2}  + \frac{9 g_R^2  }{32\pi^2}   \delta_{\pi}  
\left(1 - \frac{ g_R^2 v^2_\Sigma}{2 M_\rho^2} \right)
\frac{M_\rho^2}{F^2}
-\frac{9 g_R^4}{64\pi^2} \log\left(\frac{M_\rho^2}{M^2_{W_R}}\right)  +\mathcal{O}(g_Lg_R,g_L^2) \, .  
\ee
In order to satisfy the physical conditions in Eq. \eqref{phcond} for $c_2$, one needs $M_T \approx 2.5 F$, neglecting the gauge contributions which here enter at higher order in the insertions of explicit symmetry breaking terms (i.e. $\delta_\pi <<1$). For $c_1$, the first term on the RHS in Eq. \eqref{c1eq} can be discarded by imposing additional symmetries in the fermion mass terms forcing a cancellation between left-handed and right-handed contributions. Two additional ingredients are needed:
\begin{itemize}
\item{} A relatively large $g_R \equiv g_{R,3}$,  still within the perturbative regime, such that the gauge contribution reaches the same size as the fermion one. 
More precisely, we need $g_R^2 \times M^2_\rho/(6F)^2 \gtrsim 1$, that for natural values of $M_\rho$ implies $g_{R,3} =O(1) \gg g_{R,12} \approx g_Y^{\rm SM}$.  
\item{} A relatively light $M^2_{W_R}$, so as not to suppress (or even change the sign of) the gauge contribution.
As can be inferred from Eq. \eqref{c1eq}, the condition required is $M^2_{W_R}  = \frac{1}{4} g_R^2 \langle \Sigma_R \rangle^2 < \frac{1}{2} M^2_\rho$.
\end{itemize}
These two conditions illustrate the interesting and predictive interplay between the two main model building hypotheses, namely flavour non-universality and Higgs compositeness: the possibility to minimize the tuning in the Higgs potential directly benefits from the assumption of flavour non-universality (i.e. $g_{R,3}$ is allowed to be of $\mathcal{O}(1)$) while the impact of the scalar link fields on the potential anchors the scale of the flavour deconstruction to lie within the few TeV domain, thereby implying interesting phenomenology related to the massive flavoured gauge bosons.
\section{Phenomenology}
The phenomenological constraints are related to the two main structural features of the model: the composite dynamics, responsible for the (model-independent) modifications of Higgs couplings to SM fields, and the presence of flavoured gauge bosons, which, when integrated out at the TeV scale, affects various precision flavour and electroweak observables, as well as the high-$p_T$ distributions of $pp\to \ell\ell$ LHC measurements:
\begin{itemize}
\item \underline{Composite Dynamics}: Higgs coupling modifications and direct searches of (composite) new degrees of freedom~\cite{ATLAS:2020qdt,CMS:2020gsy} set limits on the value of $\xi = \frac{v_{\rm EW}^2}{4F^2}$ as well as on the masses $M_T$ and $M_\rho$ of the top partners and the composite spin-1 resonances, respectively. Corrections to electroweak precision observables are controlled by $\delta_{\rm EW} = g_{L,R}^2 \frac{v^2}{M^2_\rho}$ which have to be below the permille level in order to satisfy present constraints. Aiming at the lowest possible tuning in the Higgs potential, a reference benchmark point compatible with experimental constraints is given by: \begin{equation}
    F \approx 750 \text{~GeV}, \qquad 
    M_T\approx 1.8\div 2.0 \text{~TeV}, \qquad
 M_\rho\approx 8\div 10 \text{~TeV}\, .
\end{equation}
\item \underline{Flavoured gauge bosons}: The spontaneous breaking of the deconstructed EW sector (Eq. \eqref{decew}) yields two $Z^\prime$ and a pair of $W_R^\pm$ bosons dominantly coupled to the third family, with masses around the TeV scale. Once integrated out, the most stringent constraints~\cite{Allwicher:2023shc,Barbieri:2023qpf} on these new states come from $B \rightarrow X_s \gamma$ transitions and $Z$-pole observables (such as the measurement of $R_\tau$ at LEP), while flavour changing neutral current processes and high-$p_T$ Drell-Yan productions~\cite{ATLAS:2020zms,CMS:2021ctt} provide slightly less constraining bounds. A realistic benchmark point is given by: 
\be
\langle \Sigma_R \rangle \approx \langle \Omega_f \rangle \approx 3 \text{~TeV} \, .
\ee 
\end{itemize}

\section{Conclusion and Outlook}
We have introduced a model addressing the origin of flavour and electroweak stability by combining flavor non-universal gauge interactions with Higgs compositeness. The model predicts distinct phenomenological signatures, such as non-standard effects from flavoured gauge bosons, modified Higgs couplings, and light top partners, which could be probed at the LHC and future $e^+e^-$ colliders~\cite{Stefanek:2024kds}, while also offering interesting forward directions to explore alternative and enlarged global symmetries of the strong sector.

\section*{Acknowledgments}
MP is grateful to Lukas Allwicher, Sebastiano Covone, Joe Davighi and Gino Isidori for helpful discussions and comments on the manuscript. MP thanks the organizers of the Blois conference. MP's work is funded by the European Research Council (ERC) under the European Union’s Horizon 2020 research and innovation programme under grant agreement 833280 (FLAY), and by the Swiss National Science Foundation (SNF) under contract 200020-204428.

\end{document}